\documentclass[a4paper,twoside]{article}

\usepackage{epsfig}
\usepackage{multicol}
\usepackage{multirow}
\usepackage{pslatex}

\usepackage{url}

\usepackage{breakurl}

\usepackage{SCITEPRESS}     

\begin{document}

\title{TauBench 1.1: A Dynamic Benchmark for Graphics Rendering}

\author{\authorname{Erfan Momeni Yazdi, Markku M\"{a}kitalo, Julius Ikkala, and Pekka J\"{a}\"{a}skel\"{a}inen}
\affiliation{Tampere University, Finland}
\email{\{erfan.momeniyazdi, markku.makitalo, julius.ikkala, pekka.jaaskelainen\}@tuni.fi}
\datecustom{May 8, 2023}
}

\maketitle

\section*{Abstract}
Many graphics rendering algorithms used in both real-time games and virtual reality applications can get performance boosts by temporally reusing previous computations. 
However, algorithms based on temporal reuse are typically measured using trivial benchmarks with very limited dynamic features.
To this end, in \cite{alanko2022taubench} we presented TauBench 1.0, a benchmark designed to stress temporal reuse algorithms. Now, we release\footnote{\url{https://zenodo.org/record/7906987}} TauBench version 1.1, which improves the usability of the original benchmark. In particular, these improvements reduce the size of the dataset significantly, resulting in faster loading and rendering times, and in better compatibility with 3D software that impose strict size limits for the scenes.

\section{Background and Motivation}
In order to provide an interactive 3D experience in a virtual world, images must be displayed at a high frame rate. Although rendering a realistic image takes a considerable amount of time, the next frame is usually coherent with the previous frame~\cite{amortizedSS}. This kind of coherency can make rendering easier if it is utilized properly. These methods are known as temporal reuse methods, which refers to the use of a previously rendered image in order to accelerate the computation for rendering a new image. 

For testing and comparing these algorithms, we need benchmarks. Benchmarks consist of reproducible scenarios that are passed into algorithms as inputs. For testing temporal reuse algorithms, a dataset consisting of a high number of animated objects in a plausible 3D scene would be an ideal benchmark. Benchmarking allows us to compare the results of different algorithms fairly and to determine which algorithms perform better. 

There are, however, very few public datasets like these, and graphics research seldom uses them. In most cases, researchers create their own datasets or purchase a set of animations that cannot be released to the public. For this reason, TauBench was developed and made publicly available. TauBench is a benchmark specifically designed to test algorithms that leverage temporal reuse. It consists of two dynamic benchmarks, ``Eternal Valley FPS'' and ``Eternal Valley VR'' (Figure~\ref{fig:eter}), released with a permissive CC-BY-NC-SA 4.0 license. They have rapid camera movements and contain animations that are challenging for temporal rendering.

The first release of TauBench, however, had some problems. We will discuss these issues, as well as the fixes and improvements, in the next section. 

\section{Improvements}
\begin{figure*}[htp]
    \centering
    \includegraphics[width=0.78\linewidth]{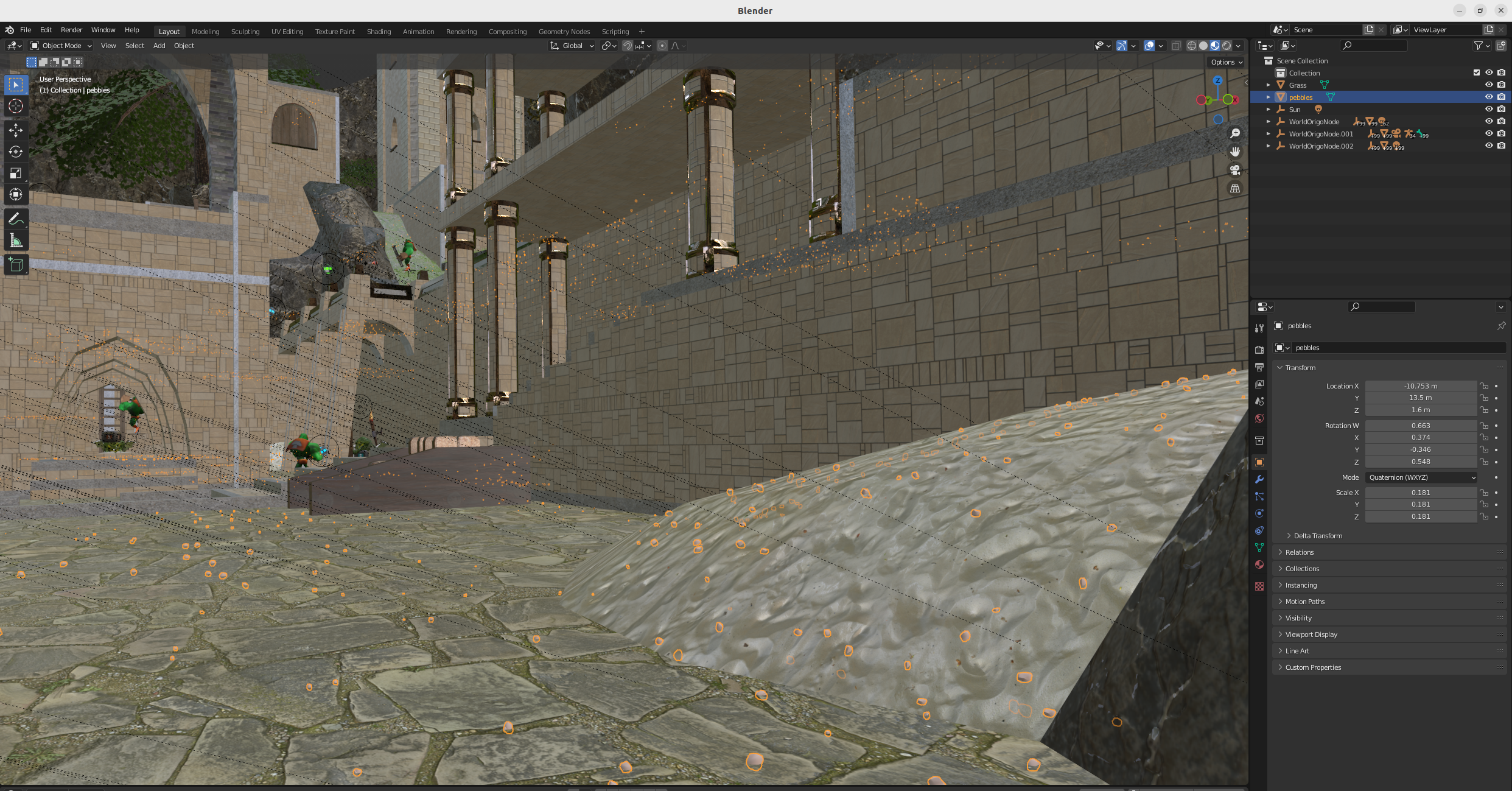} \\
    \includegraphics[width=0.78\linewidth]{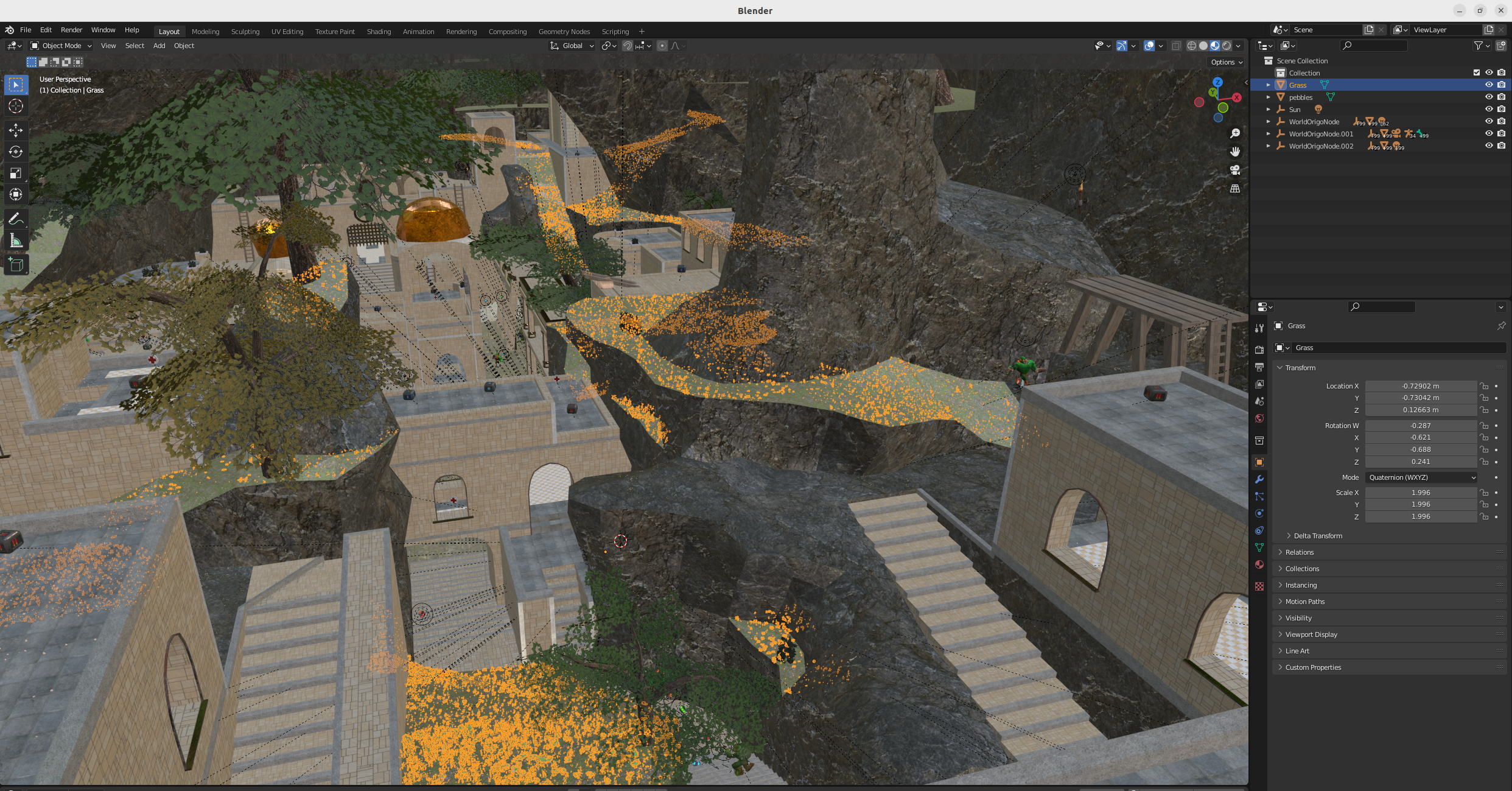}
    \caption{Pebbles (top) and Grass (bottom) in the scene were originally merged into one large mesh.}
    \label{fig:gnp}
\end{figure*}

First and foremost, version 1.0 had large file sizes, which made it difficult to load scenes in, e.g., Blender. This problem was caused by two factors. First, the pebbles were combined into a large mesh, resulting in each pebble duplicating the same source mesh data. The grass in the scenes also had the same issue. Figure~\ref{fig:gnp} shows these problems in the scene.

\begin{figure*}[htp]
    \centering
    \includegraphics[width=0.78\linewidth]{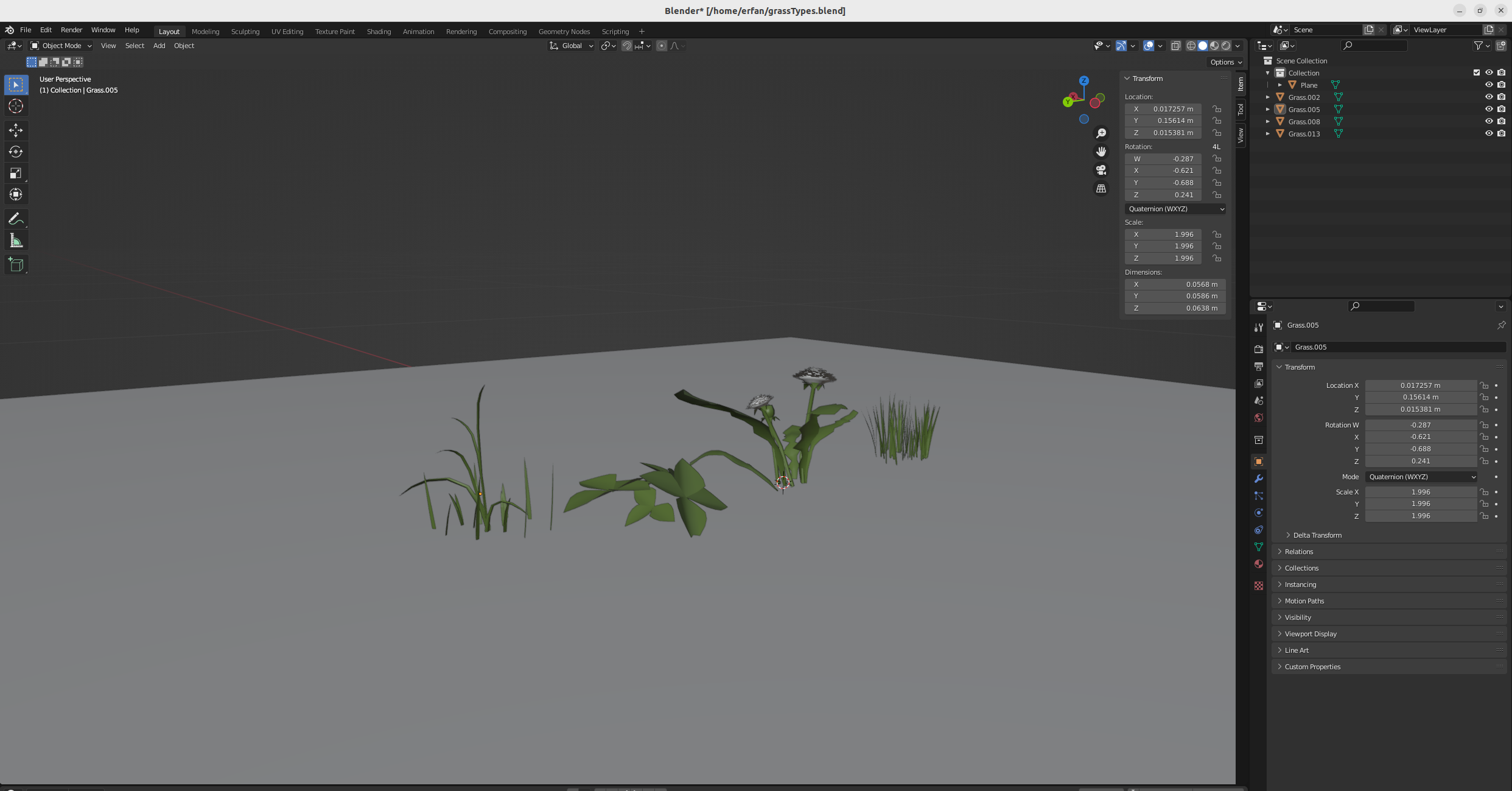}
    \caption{Different grass types separated from the original mesh. These meshes were used in Blender's particle system.}
    \label{fig:grass}
\end{figure*}

In order to fix the pebbles, we separated the unique meshes, then made instances of those meshes and scattered them where they originally were. We did the same procedure for the grass, although instead of using scripts to scatter the meshes we used Blender's particle system. Figure~\ref{fig:grass} shows different grass types that were used in the particle system. 

After these two issues were resolved, the file size of each scene was reduced from 2.7~GB to approximately 900~MB. As a result, importing the scenes in Blender became much faster: With the new version, importing the scene into Blender~3.4 takes 12 seconds instead of 36 seconds with the first version.

\begin{figure*}[htp]
    \centering
    \includegraphics[width=0.49\linewidth]{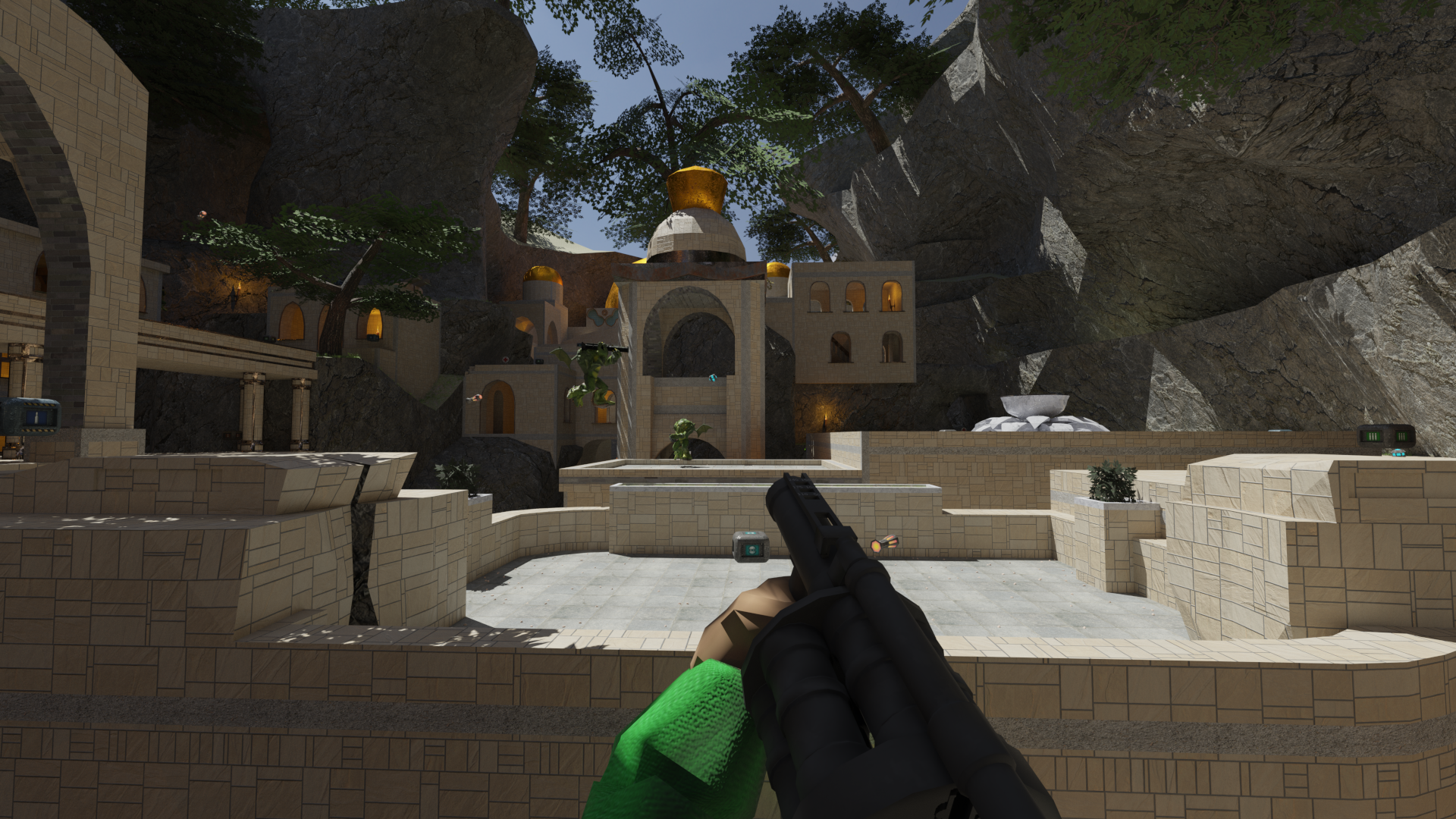}
    \includegraphics[width=0.49\linewidth]{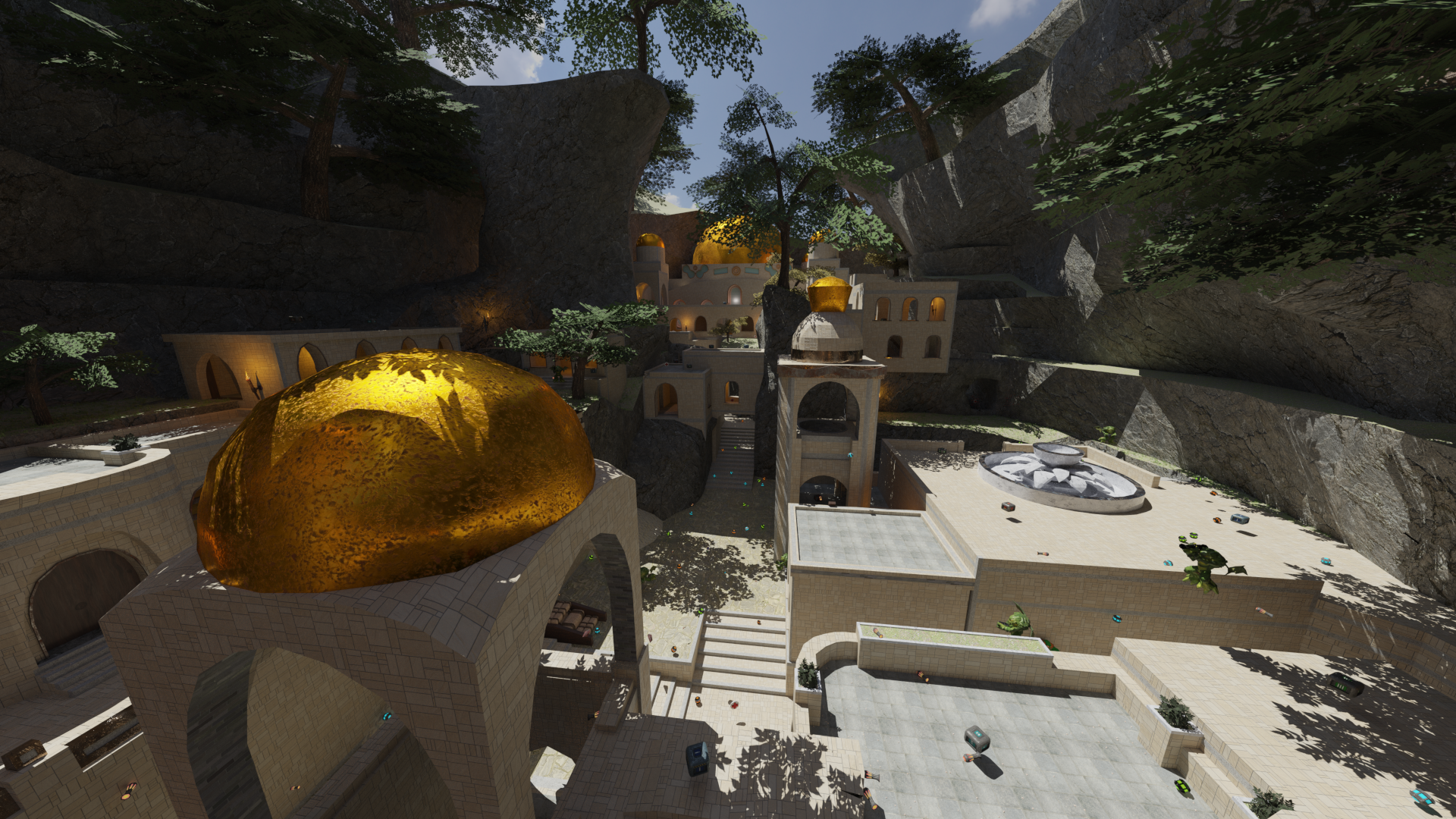}
    \caption{Eternal Valley scene (left: FPS, right: VR) rendered with Tauray at 4K resolution with 4096 spp.} 
    \label{fig:eter}
\end{figure*}

\begin{figure*}[htp]
    \centering
    \includegraphics[width=0.78\linewidth]{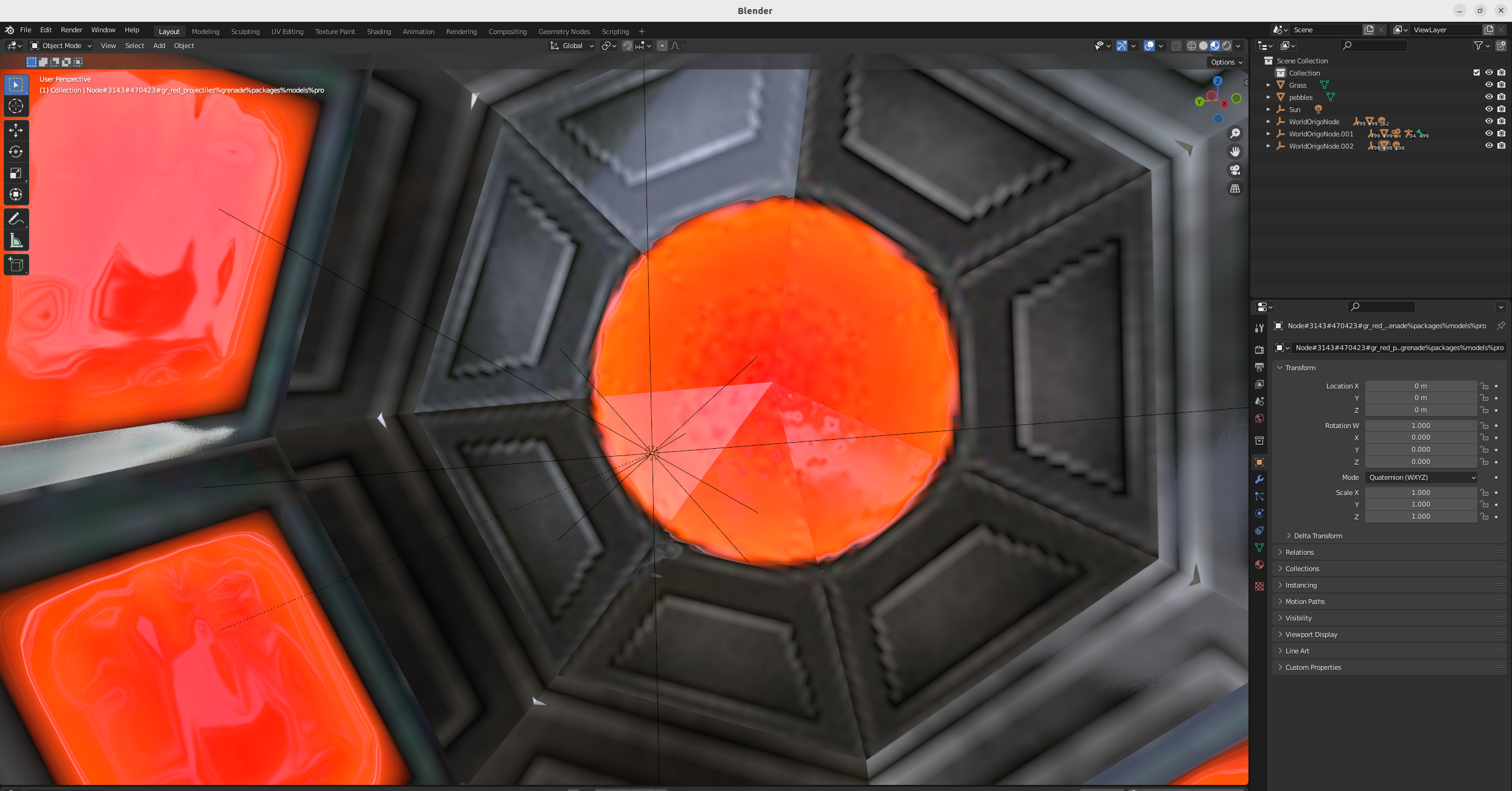}
    \caption{Point lights inside the grenades were always fully occluded in path tracers.}
    \label{fig:grenade}
\end{figure*}

Another issue was related to the dynamic point lights. Specifically, there are many point lights inside moving grenades in the original scene. However, since these lights are occluded by geometry, they have no effect on the scene with accurate shadow maps or ray-tracing based methods. Figure~\ref{fig:grenade} illustrates this problem. As a result, the ray-tracing performance and quality were negatively affected, so we decided to remove the point lights inside the grenades completely. This reduced the rendering time significantly.

We evaluated the effect of the above changes by path tracing the scenes using the Tauray renderer~\cite{Ikkala22}, the Falcor renderer~\cite{falcor}, and Blender's~\cite{Blender} Cycles renderer, on a PC with an Intel's 12th Gen Core™ i7-12700 CPU and an NVIDIA RTX 3080 GPU.

As a result of fixing the issues, with Tauray, the average frame rendering time was reduced from 27~ms to 13~ms at 1080p resolution, with one sample per pixel (spp). Also with 4096~spp at 4K resolution, the average rendering time was reduced from 260~s to 170~s. Figure~\ref{fig:eter} shows two example images rendered with Tauray.

\begin{figure*}[htp]
    \centering
    \includegraphics[width=0.78\linewidth]{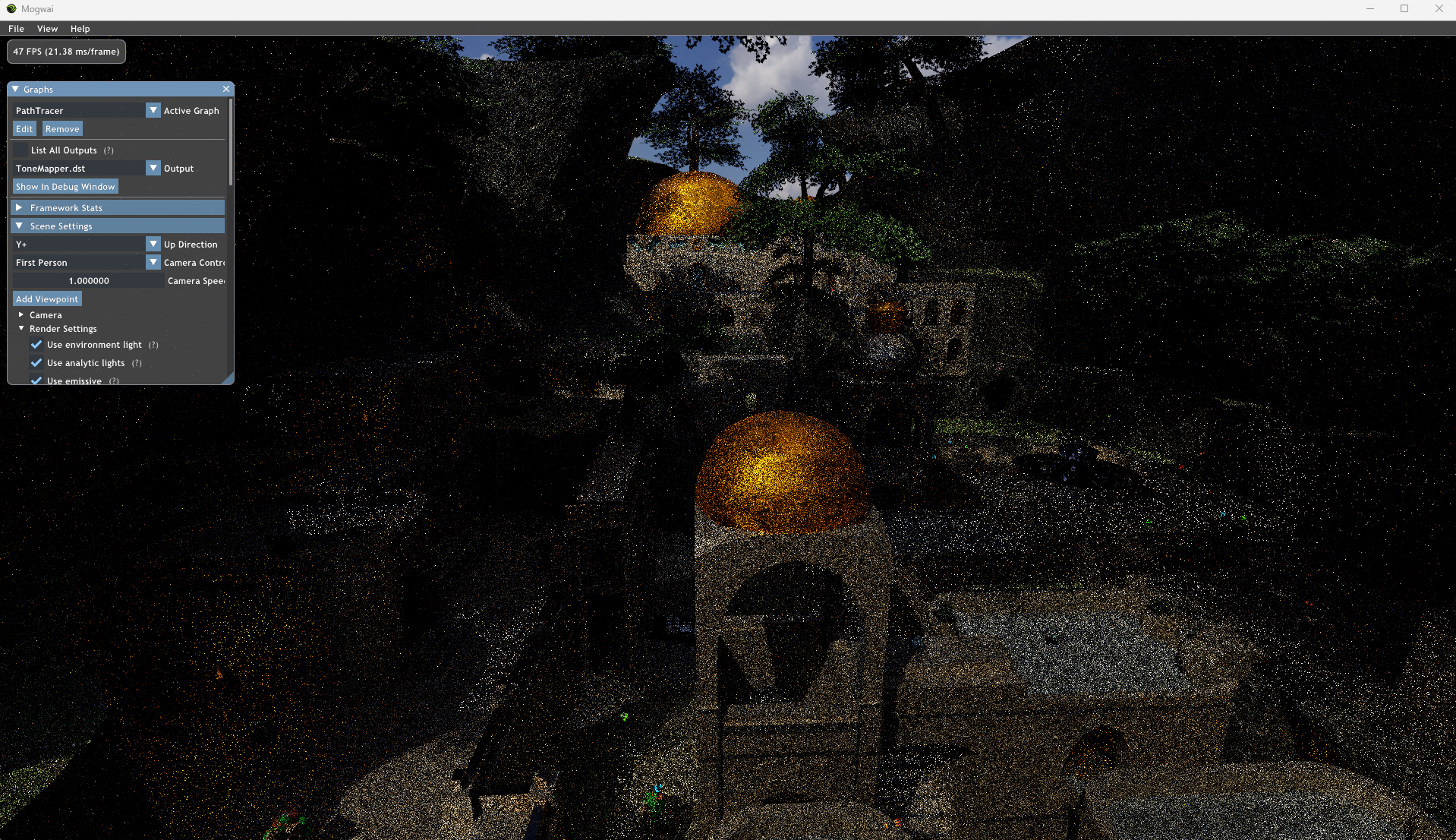} \\
    \vspace{0.1cm}
    \includegraphics[width=0.78\linewidth]{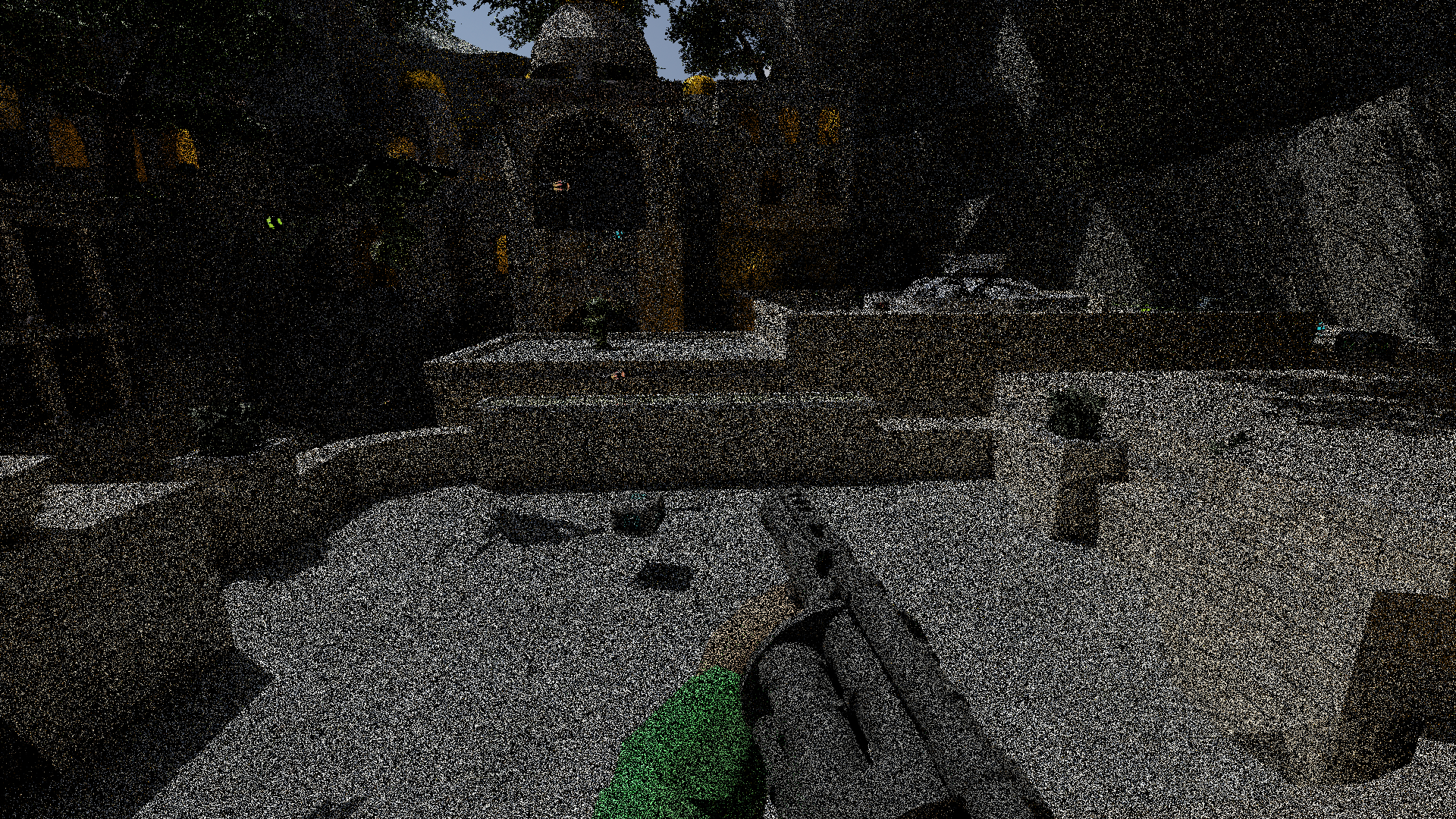}
    \caption {Falcor (top) and Tauray (bottom) renderers, rendering TauBench 1.1 with 1 spp at 1080p resolution.}
    \label{fig:falcor}
\end{figure*}

With Falcor, we could not load the original scenes due to their large file size. However, with the new version of TauBench, the average frame time was 21~ms at 1080p resolution and 1~spp. Figure~\ref{fig:falcor} shows example 1~spp renders produced by Falcor and Tauray.

Additionally, we rendered the scenes in Blender. As the Cycles renderer is not targeted towards real-time path tracing, we only tested offline rendering. With 4096~spp at 4K resolution, the average rendering time was reduced from 1332~s to 900~s.

\section{Temporal Benchmarking}
TauBench is targeted towards rendering algorithms that reuse temporal data. Hence, in this section, we introduce a set of practical benchmark targets in order to make it easier to compare various algorithms.

In practice, we fix a set of PSNR quality targets: 18~dB, 20~dB, 22~dB, 24~dB, 26~dB, and 28~dB. We consider a target achieved for a scene, when the weighted average PSNR of all the rendered frames reaches the target value. Specifically, the weighting is done by ignoring the 10 best and 10 worst frames (in terms of PSNR), and computing the average PSNR of the remaining frames. This relaxation allows for, e.g., initial temporal data accumulation during the first frames without it affecting the results.

We evaluate the proposed targets for 1) path traced data without any temporal reuse, 2) path traced data with TAA~\cite{karis2014high}, and 3) path traced data with SVGF~\cite{schied2017spatiotemporal}, all with Tauray\footnote{\url{https://github.com/vga-group/tauray}, commit 57edb6c} and its implementations of the respective algorithms. The results are presented in Table~\ref{tab:tauray-benchmarks}, with the PSNR values computed against 1080p reference frames path traced with Tauray at 16384~spp (gamma-corrected and clipped to $[0, 255]$).

The path traced results serve as a comparison point, showing the \emph{effective spp} obtained with either TAA or SVGF for each target PSNR: for example, EternalValleyFPS with 4~spp path tracing and TAA effectively corresponds to 22~spp path traced data. Furthermore, the TAA and SVGF results demonstrate the challenging dynamic nature of the scenes, as some of the higher PSNR targets cannot be reached with any input spp. This is due to the temporal reuse eventually decreasing the quality instead of increasing it, as the accumulation blur becomes the dominating source of error instead of the noise variance. As an illustrative example, EternalValleyFPS with 1~spp path tracing and SVGF yields an average of 22.64~dB, and 64~spp + SVGF yields 23.66~dB, but 1024~spp + SVGF brings only negligible further improvements at 23.67~dB. At very high input spp counts, the PSNR begins to decrease, with the extreme case of 16384~spp + SVGF (i.e., applying SVGF on the reference frames) yielding only 23.00~dB.

\begin{table*}
\small
\centering
\caption{Benchmark results for various rendering algorithms in Tauray, in terms of target PSNR quality. The PSNR values are averages over all frames, with the 10 best and 10 worst frame-wise PSNR values removed before averaging. The average frame-wise rendering times (\texttt{HOST} times on Tauray) on an RTX 2070 Super are also reported. N/A means the target was not reached with any input spp.}
\label{tab:tauray-benchmarks}
\begin{tabular}{p{3.2cm}||p{1.6cm}|p{1.6cm}|p{1.6cm}|p{1.6cm}|p{1.6cm}|p{1.6cm}}
    \hline
    \textbf{EternalValleyFPS} & {\textbf{$\geq$ 18~dB}} & {\textbf{$\geq$ 20~dB}} & {\textbf{$\geq$ 22~dB}} & {\textbf{$\geq$ 24~dB}} & {\textbf{$\geq$ 26~dB}} & {\textbf{$\geq$ 28~dB}}\\
    \hline
    \hline
    \textsc{Path tracing (PT)}$^a$ & 14 spp & 22 spp & 35 spp & 59 spp & 99 spp & 167 spp\\
    & 0.67 s$^d$ & 1.18 s & 1.89 s & 3.21 s & 5.41 s & 9.10 s\\
    \hline
    \textsc{PT + TAA}$^b$ & 3 spp & 4 spp & 7 spp & 13 spp & 36 spp & \multirow{2}{*}{N/A}\\
    & 0.13 s & 0.17 s & 0.31 s & 0.61 s & 1.74 s & \\
    \hline
    \textsc{PT + SVGF}$^c$ & \multicolumn{3}{|c|}{1 spp} & \multirow{2}{*}{N/A} & \multirow{2}{*}{N/A} & \multirow{2}{*}{N/A}\\
    & \multicolumn{3}{|c|}{0.064 s} & & & \\
    \hline
\end{tabular}

\bigskip

\begin{tabular}{p{3.2cm}||p{1.6cm}|p{1.6cm}|p{1.6cm}|p{1.6cm}|p{1.6cm}|p{1.6cm}}
    \hline
    \textbf{EternalValleyVR} & {\textbf{$\geq$ 18~dB}} & {\textbf{$\geq$ 20~dB}} & {\textbf{$\geq$ 22~dB}} & {\textbf{$\geq$ 24~dB}} & {\textbf{$\geq$ 26~dB}} & {\textbf{$\geq$ 28~dB}}\\
    \hline
    \hline
    \textsc{Path tracing (PT)} & 12 spp & 20 spp & 33 spp & 58 spp & 101 spp & 180 spp\\
    & 0.62 s & 1.04 s & 1.74 s & 3.07 s & 5.37 s & 9.52 s\\
    \hline
    \textsc{PT + TAA} & 2 spp & 4 spp & 6 spp & 11 spp & 22 spp & 61 spp\\
    & 0.086 s & 0.17 s & 0.27 s & 0.51 s & 1.01 s & 2.91 s\\
    \hline
    \textsc{PT + SVGF} & \multicolumn{4}{|c|}{1 spp} & 4 spp & \multirow{2}{*}{N/A}\\
    & \multicolumn{4}{|c|}{0.060 s} & 0.19 s & \\
    \hline
\end{tabular}

\bigskip

\begin{tabular}{c|p{13cm}}
\hline
& \textbf{Tauray parameters}\\
\hline
\hline
\textit{a} & \texttt{tauray EternalValley\{FPS,VR\}\_1.1.glb --renderer=path-tracer --envmap=kloofendal\_48d\_partly\_cloudy\_8k.hdr --animation --force-double-sided --camera-clip-range=0.1,10000 --validation=off --width=1920 --height=1080 --film=blackman-harris --tonemap=gamma-correction --filetype=png --headless=result --rng-seed=0 --max-ray-depth=4 --samples-per-pixel=<spp>}\\
\hline
\textit{b} & \texttt{--taa=8 --regularization=0.5} appended to \textit{a}\\
\textit{c} & \texttt{--denoiser=svgf} appended to \textit{b}\\
\hline
\textit{d} & \texttt{-t --filetype=none} appended for all timings (to skip writing the output frames on disk)
\end{tabular}
\end{table*}

\section{Conclusion}
We discussed the problems with TauBench version 1.0, as well as how we solved them in version 1.1 and how these solutions improved the performance. Nevertheless, there are still some improvements that could be made. For example, the scenes now lack dynamic lights that would further challenge temporal reuse based algorithms. Furthermore, we could have used more reflective materials. We plan to include new scenes with these features, along with more complicated materials and animations, in a later version of TauBench. We also introduced a set of quality targets in order to facilitate easier benchmarking of temporal reuse algorithms, and presented the benchmark results for TAA and SVGF.

\bibliographystyle{ieeetr}
{\small \bibliography{bibliography}}

\end{document}